\documentclass[dvips,12pt]{elsart}
\usepackage{epsf}
\usepackage[dvips]{graphicx}
\begin{document}
\begin{frontmatter}
\title{{\small\rm
\rightline{IIT-HEP-99/3}
\rightline{Fermilab-Conf-00/019}
\rightline{physics/0001037}
}
Muon Collider/Neutrino Factory: Status and Prospects\thanksref{talk}}
\thanks[talk]{Invited 
talk presented at the {\sl 7th International Conference on Instrumentation for 
Colliding-Beam Physics}, Hamamatsu, Japan,  Nov.\ 15--19, 1999.}
\author{Daniel M. Kaplan\thanksref{email}}\thanks[email]{E-mail: kaplan@fnal.gov}
\address{Illinois Institute of Technology,
Chicago, IL 60616\\
{\rm and}\\
Fermi National Accelerator Laboratory,
Batavia, IL 60510}
\author{for the Neutrino Factory and Muon Collider Collaboration}
\begin{abstract}

During the 1990s an international collaboration has been studying the
possibility of constructing and operating a high-energy high-luminosity
$\mu^+\mu^-$ collider. Such a machine could be the approach of choice to extend
our discovery reach beyond that of the LHC. More recently, a growing
collaboration is exploring the potential of a  stored-muon-beam ``neutrino
factory" to elucidate neutrino oscillations. A neutrino factory could be an
attractive stepping-stone to a muon collider. Its construction, possibly 
feasible within the coming decade, could have substantial impact on
neutrino physics.

\end{abstract}
\end{frontmatter}

\section{Introduction}

The Neutrino Factory and Muon Collider Collaboration (NFMCC)~\cite{MuColl} is  
engaged in an international R\&D project to establish the feasibility of a  
high-energy $\mu^+\mu^-$ collider and a stored-muon-beam ``neutrino factory\@."  
As a  heavy lepton, the muon offers important advantages over the electron for 
use in a high-energy collider: 
\begin{enumerate}
\item Radiative processes are highly suppressed, allowing use of 
recycling accelerators. This reduces the size and cost of the complex.
It also allows use of a storage ring, 
increasing 
luminosity by a factor $\approx10^3$ over a one-pass collider.
\item In the Standard Model and many of its extensions, use of a 
heavy lepton increases the cross section for $s$-channel Higgs production 
by a factor ${m_\mu}^2/{m_e}^2=4.3\times10^4$, opening a unique avenue 
for studying the dynamics of electroweak symmetry breaking~\cite{Bargeretal}. 
More 
generally, a sensitivity advantage may be expected in any model that 
seeks to explain mass generation~\cite{Eichten}.
\item Beam-beam interactions make high luminosity harder to achieve as the
energy of an $e^+e^-$ linear collider is increased, an effect that is 
negligible for muon colliders~\cite{Palmer-Gallardo}.
\end{enumerate}

\begin{figure}
\centerline{\epsfxsize=\textwidth\epsffile{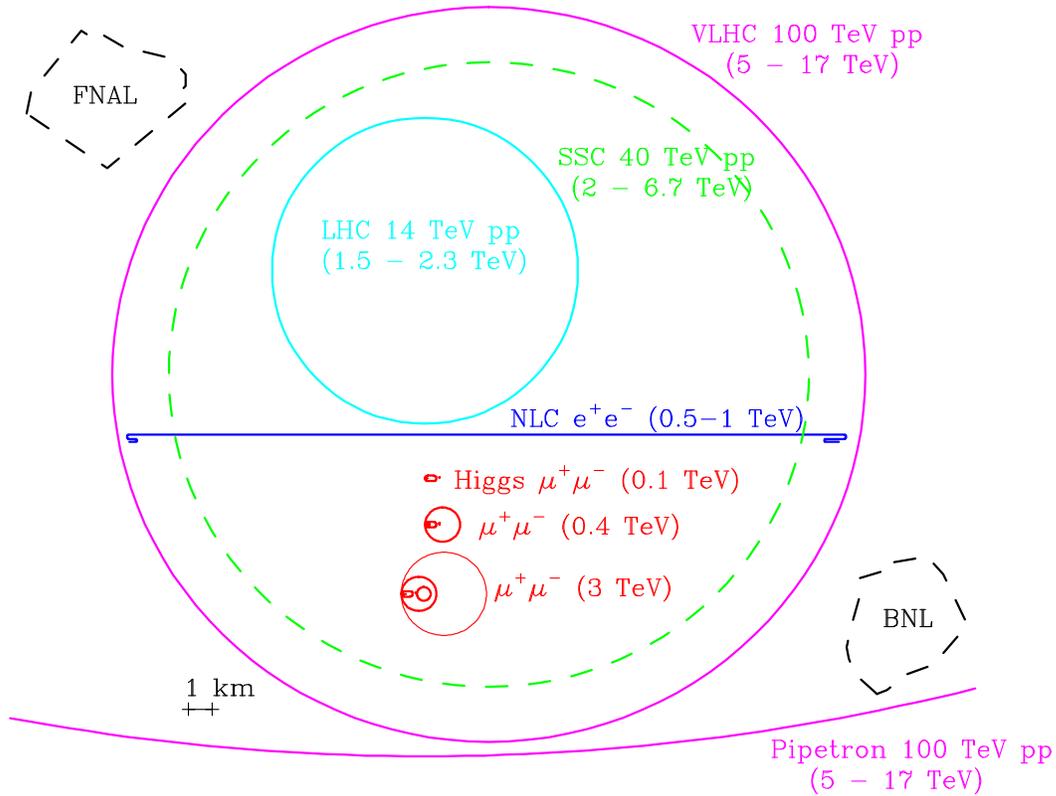}}
\caption{Sizes of various proposed high-energy colliders as compared with FNAL
and BNL sites. A muon collider with $\sqrt{s}$ up to 3\,TeV fits easily on existing
sites.\label{fig:sites}}
\end{figure}

The small size anticipated for a muon collider is indicated in 
Fig.~\ref{fig:sites}, which compares various proposed future accelerators.
Unlike other proposals, muon colliders up to $\sqrt{s}=3\,$TeV fit comfortably
on existing  Laboratory sites. Beyond this energy neutrino-induced radiation
(produced by neutrino interactions in surface rock), which increases as $E^3$,
starts to become a significant hazard, and new ideas or sites where the
neutrinos break ground in uninhabited areas would be required. A muon collider
facility can provide many ancillary benefits (physics ``spinoffs") and can be
staged to provide interesting physics opportunities even before a high-energy
collider is completed. These include experiments~\cite{Front-End} with intense
meson and muon beams produced using the high-flux ($\approx2\times10^{22}$
protons/year) proton source, as well as neutrino beams of unprecedented
intensity and quality, discussed below in Sec.~\ref{sec:nufac}.

Ref.~\cite{Status-Report} is a comprehensive summary of the status of muon
collider research as of $\approx$1 year ago. While stored muon beams have been
discussed since $\approx$1960~\cite{stored-muon}, and muon colliders since
1968~\cite{Tikhonin}, only in recent years has a practical approach to the
realization of a $\mu^+\mu^-$ collider been devised. The key concept that may
allow a muon collider to become a reality is ionization
cooling~\cite{ONeill,Lichtenberg,Neuffer2}. Muons may be copiously produced using
collisions of multi-GeV protons with a target to produce pions, which then
decay in a focusing  ``capture channel\@." However, those muons occupy a large
emittance (phase-space volume) and are unsuitable for injection directly into
an accelerator. Muon-beam cooling is needed to reduce the emittance by a
sufficient factor but must be carried out in a time short compared to the
$\approx2\,\mu$s muon lifetime. While other beam-cooling methods are  too slow,
as discussed below, simulations show that ionization cooling can meet these
requirements.

\section{Proton Source, Targetry, and Capture}

\begin{figure}
\centerline{\epsfxsize=\textwidth\epsffile{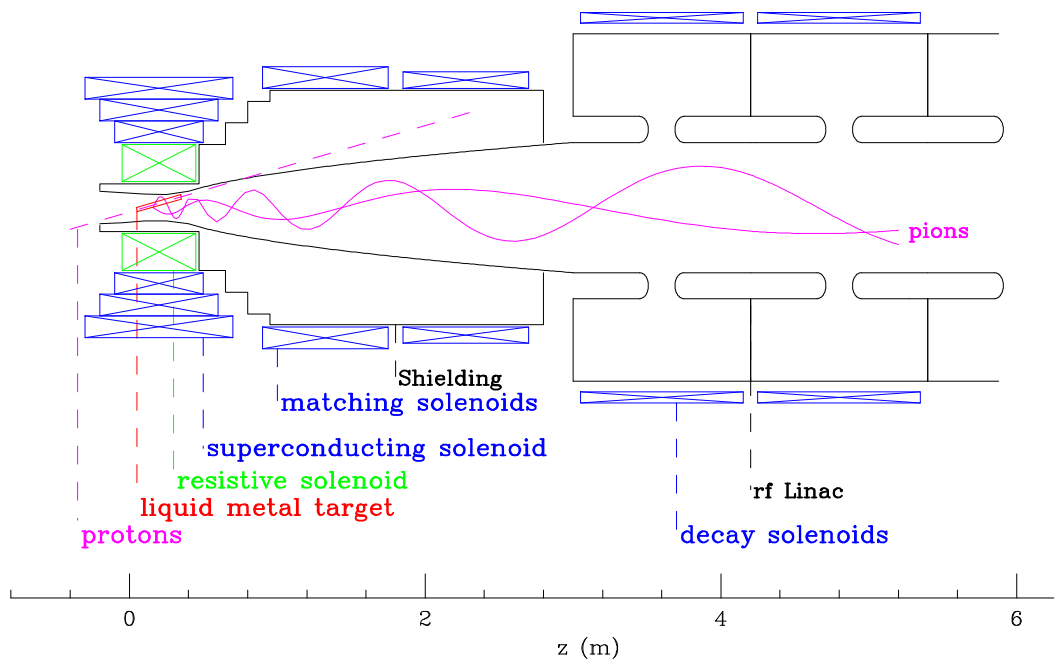}}
\caption{Liquid-jet pion-production target with solenoidal capture/decay
channel.\label{fig:targetry}}
\end{figure}

\begin{figure}
\vspace{0.5in}
\centerline{\scalebox{0.75}{\includegraphics*{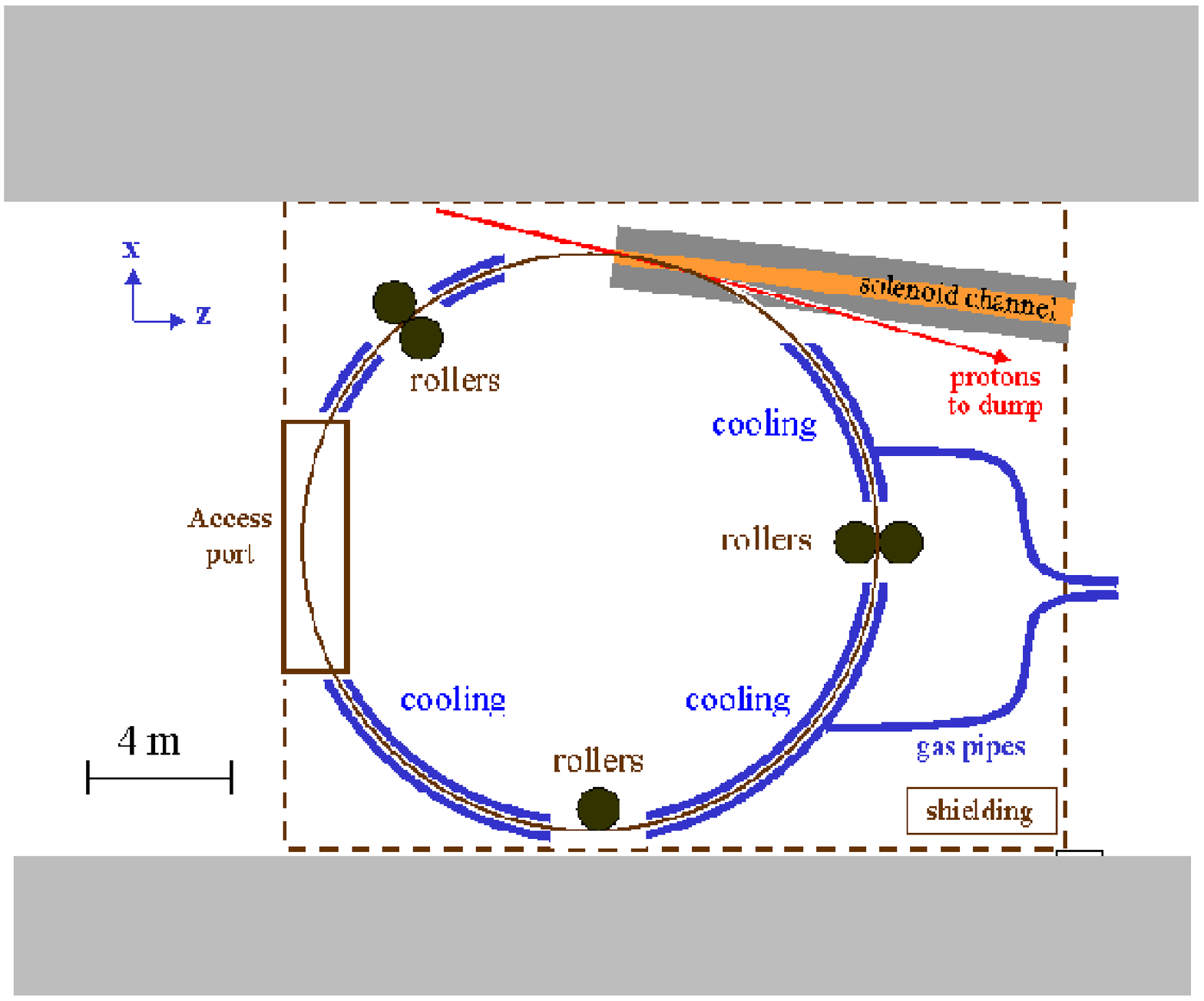}}}
\vspace{0.5in}
\caption{``Bandsaw" target concept.\label{fig:bandsaw}}
\end{figure}

Muon-collider luminosity estimates (see Table~\ref{tab:params}) have been made
under the assumption that a 4\,MW proton beam may safely strike a target
representing 2--3 hadronic interaction lengths, which is tilted at a
100--150\,mr angle with respect to the solenoidal focusing field (see
Fig.~\ref{fig:targetry}).  Design studies are ongoing to demonstrate this in
detail, and BNL-E951 at the AGS will test this experimentally within the next
few years~\cite{targetry}. Ideas being explored include a liquid-metal jet
target (Fig.~\ref{fig:targetry}) and a ``bandsaw" target
(Fig.~\ref{fig:bandsaw}). In neutrino-factory scenarios the beam power
requirement is eased to 1\,MW, making a graphite target also a possibility.
About 10\% of the beam energy is dissipated in the target.  A target with
comparable dissipation is being designed for the Spallation Neutron Source at
Oak Ridge National Laboratory~\cite{SNS}. The required 4\,MW proton source,
while beyond existing capability, is the subject of ongoing design studies at
Brookhaven~\cite{BNL-source} and Fermilab~\cite{FNAL-source} and is comparable
in many respects to machines proposed for spallation neutron
sources~\cite{BNL-SNS}. For efficient capture of the produced low-energy pions,
the target is located within a 20\,T solenoidal magnetic field to be produced
using a superconducting solenoid with a water-cooled copper-coil insert
(Fig.~\ref{fig:targetry}). The captured pions and their decay muons proceed
through a solenoidal field that decreases adiabatically to 1.25\,T. Simulations
show that $\approx$0.6 pions/proton are captured in such a channel for proton
energy in the range 16--24\,GeV.

\begin{figure}
\centerline{\rotatebox{90}{\epsfxsize=4in\epsffile{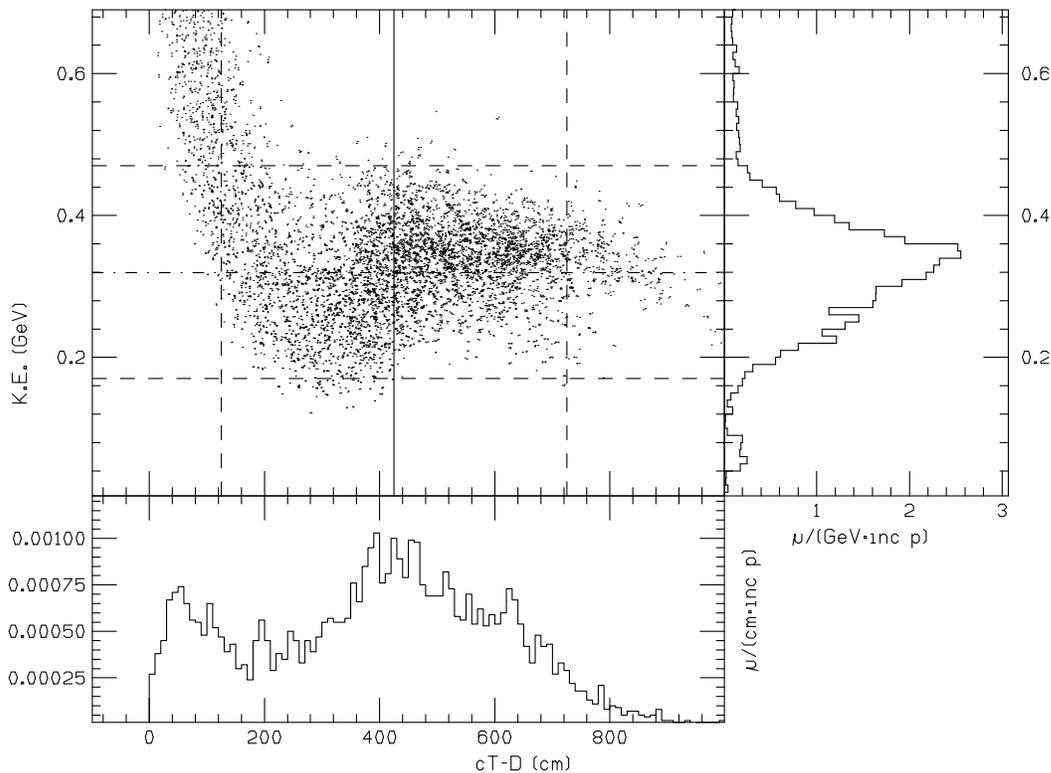}}}
\caption{Event distributions vs.\ kinetic energy and longitudinal distance 
after phase rotation.\label{fig:Evst}}
\end{figure}

The resulting muon bunches, while very intense, feature a large energy spread,
which must be decreased for acceptance into a cooling channel. This can be
accomplished via radio-frequency (RF) ``phase rotation," by which the energy of
the low-energy muons is raised and that of the high-energy muons lowered.  This
brings a substantial fraction ($\approx$60\%) of the muons into a narrow energy
range at the expense of increasing the bunch length (Fig.~\ref{fig:Evst}). The
large transverse size of the beam at this point necessitates low-frequency 
(30--60\,MHz) RF cavities. An alternative under investigation is use of an
induction linac. If the phase-rotation accelerating gradient is sufficiently
high ($\approx$4--5\,MV/m), 
a significant portion of the pions can be phase-rotated before they decay,
allowing muon polarization as high as $\approx$50\%. Otherwise the polarization
is naturally $\approx$20\%~\cite{Blondel}. Muon polarization can be exploited in a variety of
physics studies~\cite{Status-Report} as well as in a neutrino
factory~\cite{Geer}. It can
also provide a $\sim10^{-6}$ fill-to-fill relative calibration of
the beam energy~\cite{Raja} (needed e.g.\ to measure the width
of the Higgs).

\section{Ionization Cooling}

\begin{figure}
\centerline{\epsfxsize=4in\epsffile{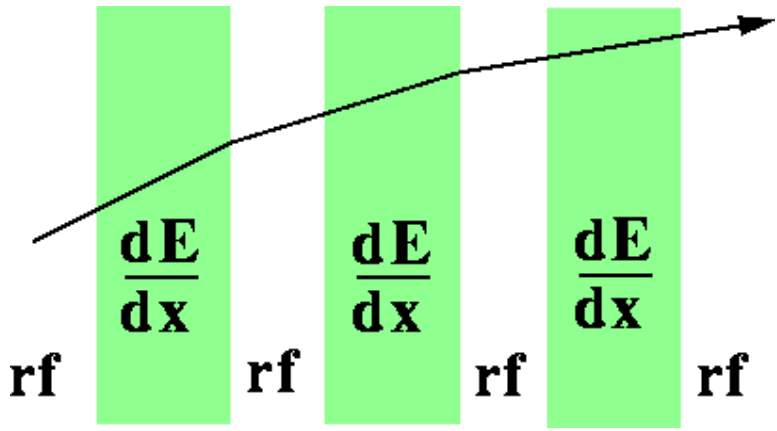}}
\caption{Principle of ionization cooling.\label{fig1}}
\end{figure}

The goal of beam cooling is to reduce the normalized six-dimensional emittance
of the beam. The significance of emittance for accelerator design is that it
places limits on how tightly the beam can be focused and determines the
apertures necessary to transport and accelerate it without losses. A muon beam
can be cooled by passing it alternately through  material, in which ionization
energy loss reduces both the transverse and  longitudinal momentum  components,
and RF accelerating cavities, in which the lost longitudinal  momentum is
restored (Fig.~\ref{fig1}). This process reduces the transverse  momentum
components relative to the longitudinal one (``transverse cooling").  

Before summarizing the theory of ionization cooling we must first define
emittance more carefully.

\subsection{Emittance}

Emittance 
can be defined in terms of the vector of canonical variables describing each
muon, which can be chosen as $\mathbf{X}=(x,y,z,p_x,p_y,p_z)$. 
In the absence of correlations among these variables, the six-dimensional volume
of the beam in phase space can be represented by 
$\sigma_x\sigma_{p_x}\sigma_y\sigma_{p_y}\sigma_z\sigma_{p_z}$, where $\sigma_i$
designates the r.m.s.\ width of the distribution in the $i$th variable.
More generally, the
normalized six-dimensional emittance $\epsilon_{6,n}$ is given by
$\sqrt{\det{V}}/(m_{\mu}c)^6$, where $V$ is the $6\times6$ covariance matrix of
$\mathbf{X}$, $m_\mu$ is the mass of of the muon, and $c$ is the speed of
light. If the off-diagonal (correlation) terms of $V$ are negligible, 
the emittance can be approximated as
$\epsilon_{6,n}\approx\epsilon_{x,n}\epsilon_{y,n}\epsilon_{z,n}$,
where $\epsilon_{x,n}=\sigma_x\sigma_{p_x}/m_\mu c$ and so forth, or in a
cylindrically-symmetric system such as we consider below, 
$\epsilon_{6,n}\approx\epsilon_{\perp n}^2\epsilon_{\|n}$, where
$\epsilon_{\perp n}$ ($\epsilon_{\|n}$) is the normalized transverse 
(longitudinal) emittance.

By Liouville's theorem, normalized emittance is conserved in linear beam 
transport and acceleration. Beam cooling thus requires a ``violation" of
Liouville's theorem, which is possible by means of dissipative forces such as
ionization energy loss~\cite{Lichtenberg}.

\subsection{Ionization-cooling theory}

Ionization cooling is approximately described by the following 
equation~\cite{Neuffer2,Fernow}: 
\begin{equation}   
\frac{d\epsilon_{\perp n}}{ds} =
-\frac{1}{\beta^2}\frac{dE_\mu}{ds} \frac{\epsilon_n}{E_\mu} + 
\frac{1}{\beta^3}\frac{\beta_\perp(0.014\,{\rm GeV})^2}{2E_\mu m_\mu L_R}\,,
\label{eq:cool}   
\end{equation}   
where $s$ is the path 
length, $E_\mu$ the muon energy, $L_R$ the radiation 
length of the absorber medium, $\beta=v/c$, and $\beta_\perp$ is the 
betatron function of the beam (inversely proportional to the square of the beam 
divergence). 

In Eq.~\ref{eq:cool} we see, in addition to the $dE/ds$ transverse cooling
term, a transverse heating term due to multiple Coulomb scattering of the muons
in the absorber. Since cooling ceases once the heating and cooling terms are
equal, Eq.~\ref{eq:cool} implies an equilibrium emittance, which in principle
(neglecting other limiting efects) would be reached asymptotically were  the
cooling channel continued indefinitely. Since the heating term is proportional
to $\beta_\perp$ and inversely proportional to the radiation length of the
absorber medium, the goal of achieving as small an equilibrium emittance as
possible requires us to locate the absorber only in low-$\beta_\perp$ regions
and to use a medium with the longest possible radiation length, namely
hydrogen. To achieve low $\beta_\perp$, we want the strongest possible focusing
elements. We are thus led to superconducting solenoids filled with liquid
hydrogen as possibly the optimal solution.\footnote{However, lithium lenses
might give an even lower equilibrium emittance than solenoids with liquid
hydrogen, since stronger focusing fields may be feasible with liquid-lithium
lenses than with magnets, and this may overcome the radiation-length advantage
of hydrogen.}

Below the ionization minimum, energy loss increases approximately as $p^{-1.7}$,
while Coulomb scattering increases only as $(p\beta)^{-1}$~\cite{PDG}.
Ionization cooling thus favors low momenta, despite the relativistic increase
in muon lifetime with momentum.  In fact, Eq.~\ref{eq:cool} implies that the
equilibrium emittance scales approximately as $\gamma^{1.7}\beta^{0.7}$. Most
simulations of muon cooling are now being done at $p=187\,$MeV/$c$. Still lower
momenta could in principle be better but are difficult to transport in practice
due to larger beam divergences.

\subsection{Cooling channel designs}

\begin{figure}
\centerline{\scalebox{.5}{\rotatebox{90}{\includegraphics{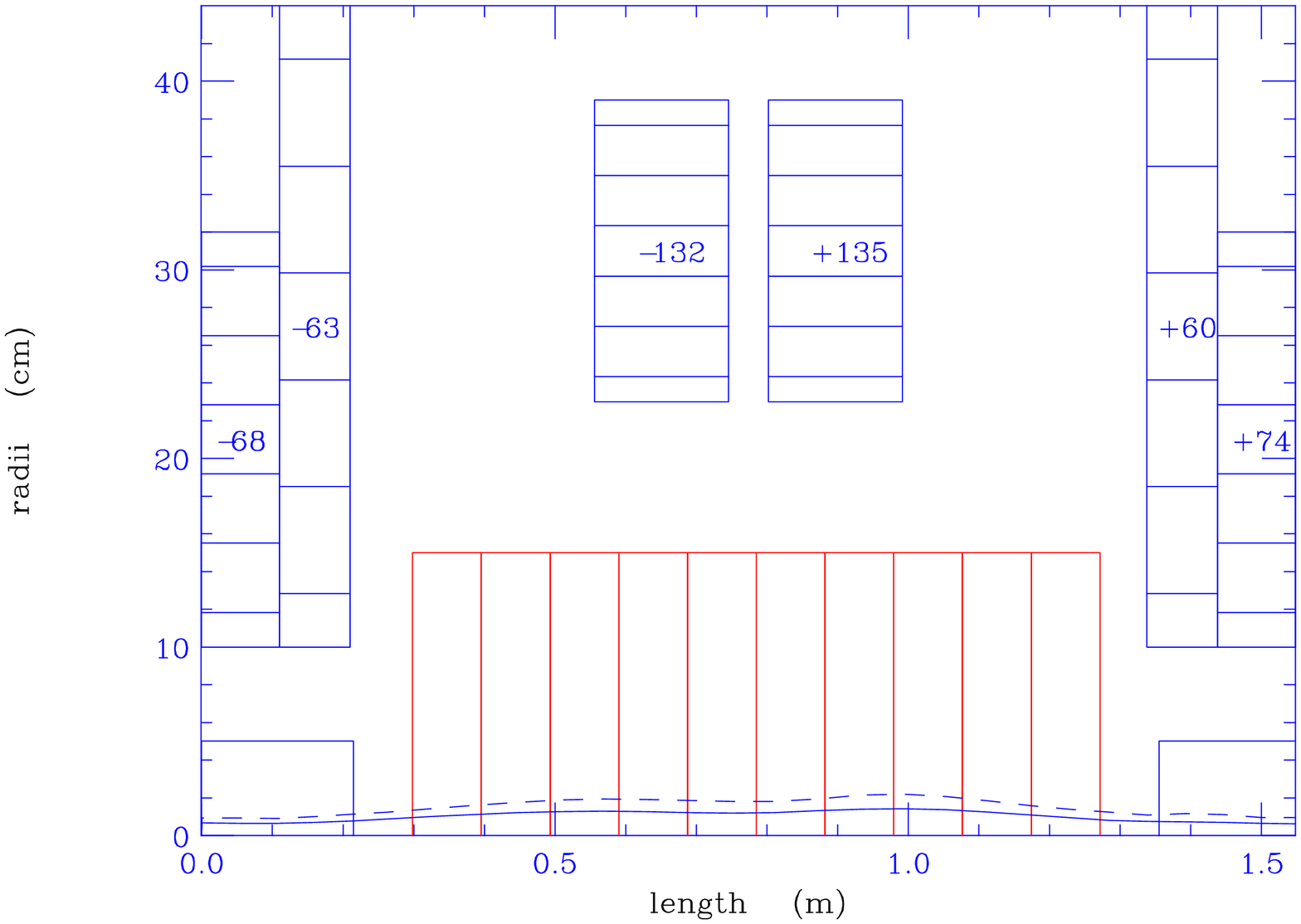}}}}
\centerline{\hspace{0.1in}\scalebox{.5}{\rotatebox{90}{\includegraphics{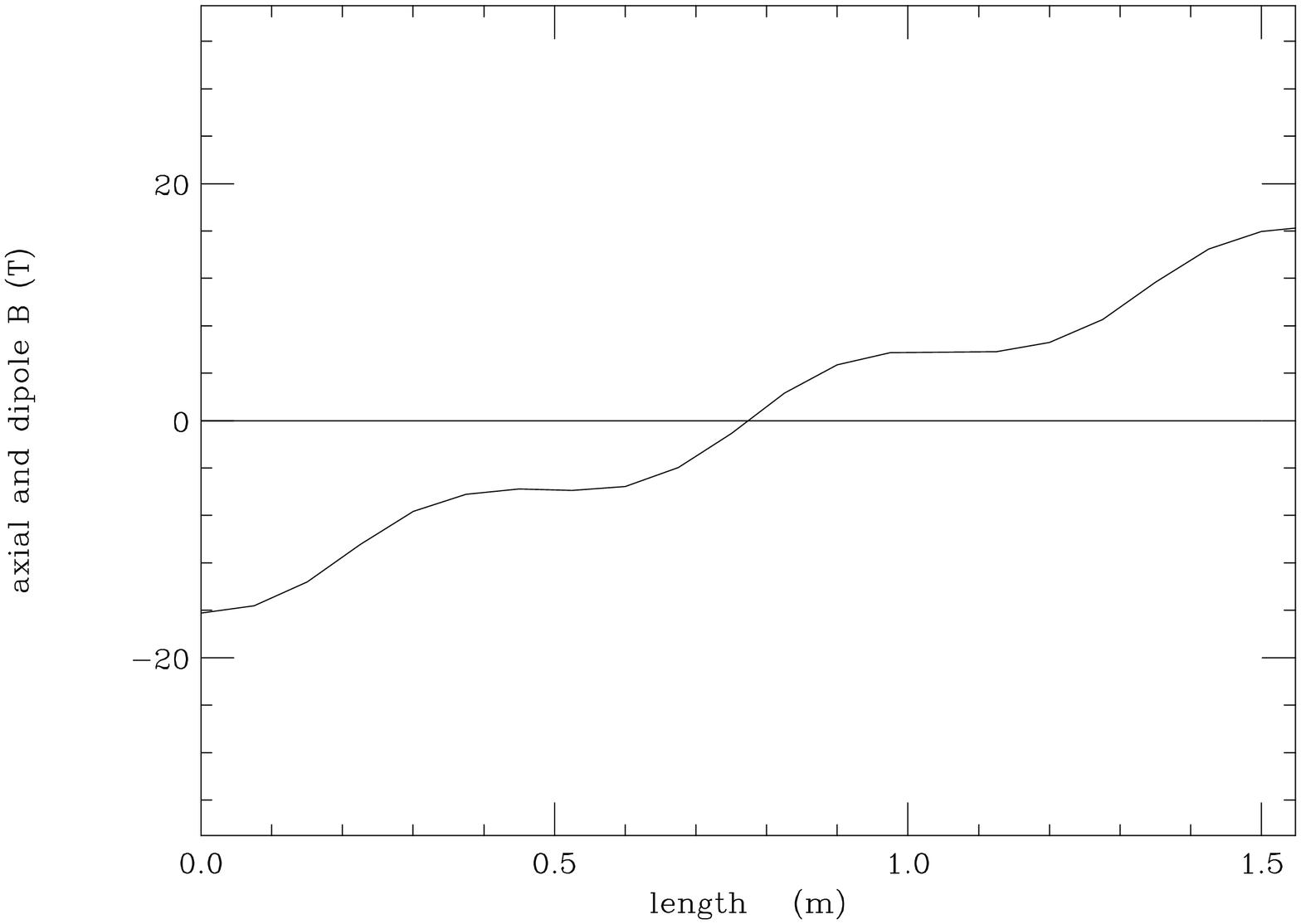}}}}
\caption[Coil configuration and field shape for 1.548\,m alternating-solenoid
lattice.]{(Top plot) coil configuration and (bottom plot) field shape on axis
for 1.548\,m alternating-solenoid lattice. Numbers shown inside coil blocks are
current densities in amperes/mm$^2$. The configuration has mirror symmetry 
about the planes at 0 and 1.548\,m such that one-half of each of two absorbers 
and high-field coils are shown. Also indicated on top plot are RF cells and
(solid and dashed curves) typical beam envelopes.
\label{fig:SoSo}}
\end{figure}

In the ``alternating-solenoid" cooling channel~\cite{2mSoSo} the muon beam is
kept focused by a series of superconducting solenoids alternating in
magnetic-field direction (Fig.~\ref{fig:SoSo}). As the field alternates,
its magnitude must of course pass through zero; at these points $\beta_\perp$
is necessarily large. Here solenoids with large inner bore are suitable,
allowing insertion of RF cavities. In between are regions of low $\beta_\perp$,
corresponding to maxima of the magnetic field, where the liquid-hydrogen
absorbers are located. (The field directions alternate so that canonical
angular momentum~\cite{Fernowetal}, which builds up within each absorber as the
muon beam loses mechanical angular momentum, cancels rather than building
up\@.)  Another type of arrangement (dubbed ``FoFo") is also under study, in
which solenoid  fringe-field focusing is employed. This allows lower
$\beta_\perp$ values for a given field strength than in the 
alternating-solenoid arrangement. In the FoFo channel the low-$\beta_\perp$
regions (and the absorbers) are thus at low field and the RF cavities at high
field. 

Both the FoFo and alternating-solenoid cooling channels feature liquid-hydrogen
absorbers in which a substantial amount of power is dissipated by the muon
beam. In a typical case, $10^{13}$ muons per bunch at a 15\,Hz repetition rate
deposit a few hundred watts in each absorber. This is within the range of
operation of high-power liquid-hydrogen targets that have been used in the
past~\cite{Mark-Beise} or proposed for future
experiments~\cite{Margaziotis-E158}. Careful attention must be paid to the
design of these absorbers, for reasons both of safety and of performance. For
example, the windows must be made of low-$Z$ material and kept as thin as
possible in order not to degrade the cooling performance by causing excessive
multiple scattering. Aluminum alloy appears to be an acceptable solution.

\begin{figure}
\centerline{\hspace{0.25in}\epsfysize=\textwidth\epsffile{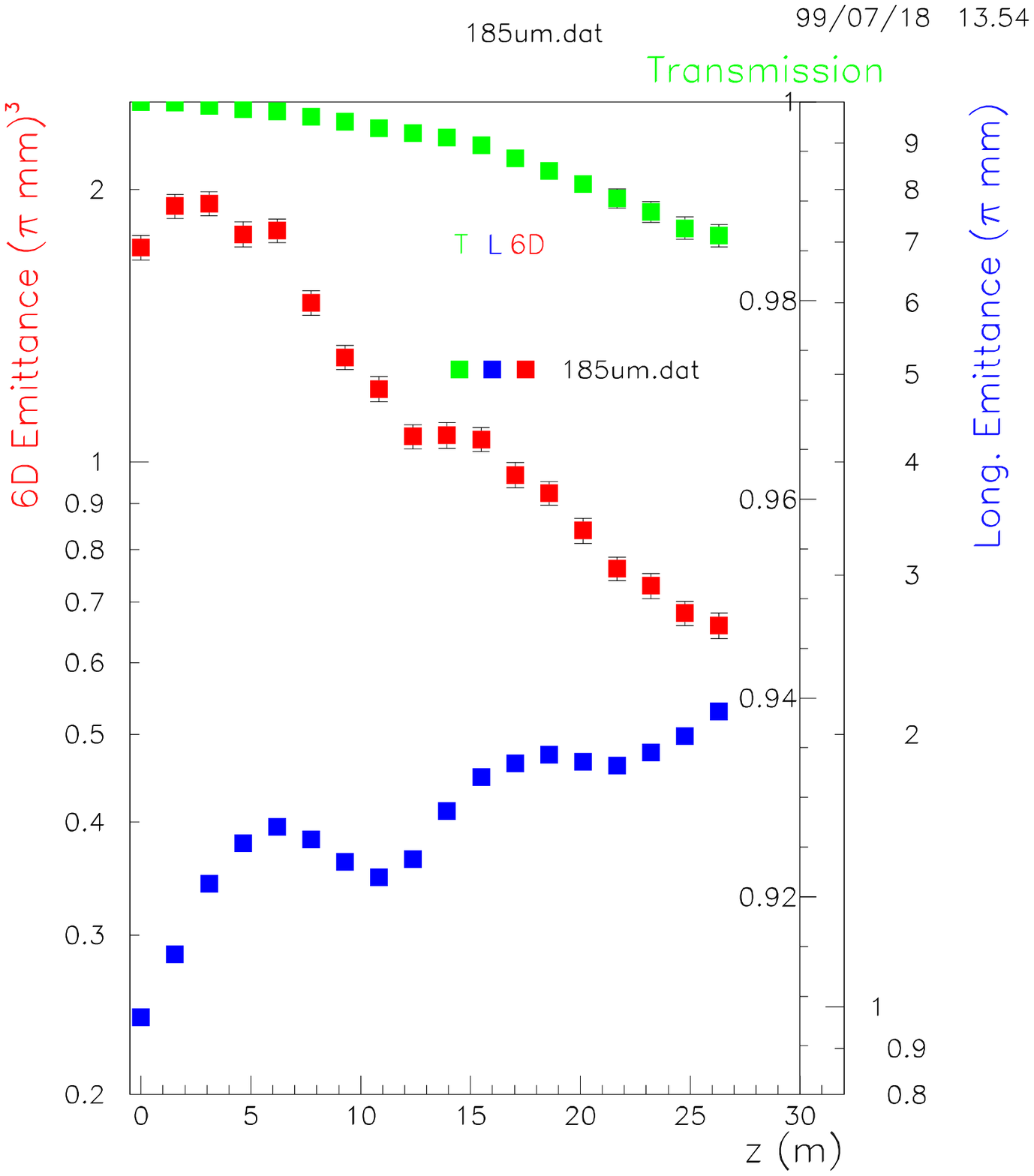}}
\caption{Results of a Geant simulation of the 1.548\,m 
alternating-solenoid
cooling lattice: (top) beam transmission (including 
non-decay losses only), 
(middle) six-dimensional emittance,
and (bottom) longitudinal emittance, all vs.\ distance.\label{fig:SoSosim}}
\end{figure}

As an example of the performance that can be achieved in such a cooling
channel, Fig.~\ref{fig:SoSosim} shows as a function of distance the
six-dimensional and longitudinal beam emittances as well as the relative beam
intensity in a simulated alternating-solenoid channel using $\approx$15\,T solenoids and
805\,MHz RF cavities. These cooling-channel parameters are representative of a
late stage of cooling for a muon collider and were chosen for initial detailed
studies in order to demonstrate a solution in a technically-challenging regime.
The six-dimensional emittance is reduced by a factor of 3 in 26\,m,  with less
than 2\% non-decay beam loss. 


Eq.~\ref{eq:cool} implies a natural scaling of the cooling-channel components
as the beam becomes progressively cooler: to maintain the  cooling rate as
equilibrium is approached, $\beta_\perp$ must be  periodically decreased to
establish a new, smaller, equilibrium emittance.  This means the focusing
fields must become stronger. How small a  $\beta_\perp$ can be achieved in an 
alternating-solenoid channel has not  yet been definitively determined, but
20\,T seems a practical upper limit  to superconducting-solenoid field
strength~\cite{Scanlan:1996qp}, and perhaps 30\,T in a 
superconducting/copper hybrid design. To continue transverse cooling  beyond
the practical limit for the alternating-solenoid channel, liquid-lithium lenses
or FoFo-type channels may be solutions.  In a scenario sketched by
Palmer~\cite{Snowmass,Status-Report} a factor $10^{6}$ in six-dimensional
emittance is achieved in a distance of $\approx$500\,m using a  series of 25 alternating-solenoid channels followed
by three lithium-lens stages, with each stage contributing a factor of about 2.
This cooling factor is sufficient to permit the collider luminosities of Table~\ref{tab:params}.

\subsection{Emittance exchange}

As the muon beam passes through the transverse-cooling channel the 
longitudinal emittance grows. This arises from four effects: 
\begin{enumerate} 
\item
Working below the ionization minimum, there is positive feedback, since as the 
muons lose momentum their energy-loss rate increases. 
\item The beam energy
spread increases in the absorber due to energy-loss straggling. 
\item  The
bunch tends to drift apart because slow muons take longer  to traverse the 
channnel than fast muons. 
\item  The bunch tends to drift apart because muons at large
transverse amplitude follow helical trajectories of greater path length than
muons at small transverse amplitude. 
\end{enumerate}

\begin{figure}
\centerline{\scalebox{0.65}{\includegraphics[clip=true]{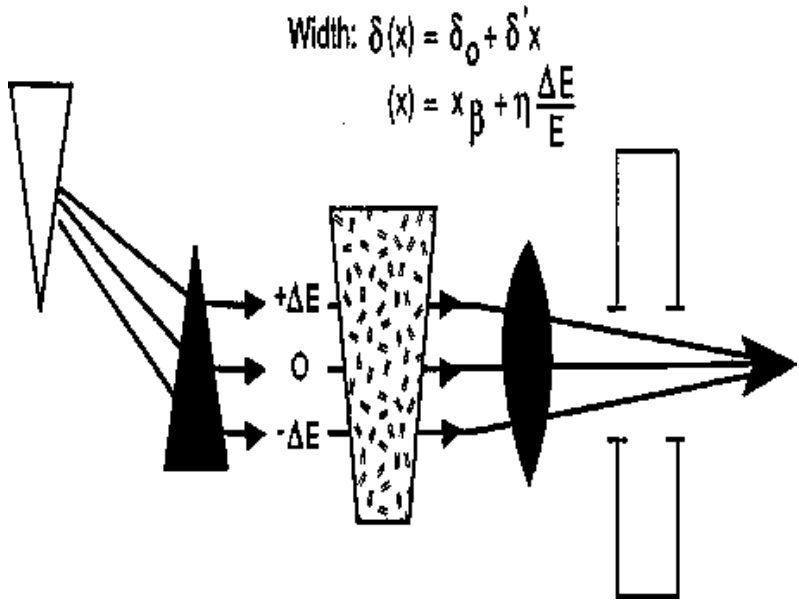}}}
\caption{Concept sketch of emittance-exchange section.\label{fig:exchange}}
\end{figure}

Eventually, significant beam losses begin to occur (see Fig.~\ref{fig:SoSosim})
as muons drift outside  the stable RF bucket. At this point longitudinal
emittance must be exchanged for transverse emittance. Such emittance exchange
may be accomplished by placing wedge-shaped absorbers  at a point of momentum
dispersion in the beam transport lattice (Fig.~\ref{fig:exchange}). With
high-momentum muons passing through a greater thickness of absorber than
low-energy muons, the beam energy spread is reduced, at the expense of an
increase in transverse beam size. While solutions have been devised on paper,
detailed simulations so far have shown that a concrete realization of this idea
is challenging, and work is ongoing to find a practical solution.

\subsection{RF Development}

Rapid muon cooling requires development of high-gradient RF cavities (e.g.,
36\,MV/m at 805\,MHz) with suitable high-power drive systems. A novel feature
of muon acceleration is the possibility of using ``pillbox" cavities,  closed
at each end by low-mass metal windows, to increase the uniformity of the
electric field within the cavity and lower the power requirements.  These
developments are in progress, with 805\,MHz~\cite{Corlett} and 175--200\,MHz
designs now under development and planned for testing over the next few years
at BNL, FNAL, and LBNL. Alternative gridded and windowless designs are also in
progress, as well as power-source design studies.

\subsection{Muon Cooling Experiment}

\begin{figure}
\centerline{\epsfxsize=\textwidth\epsffile{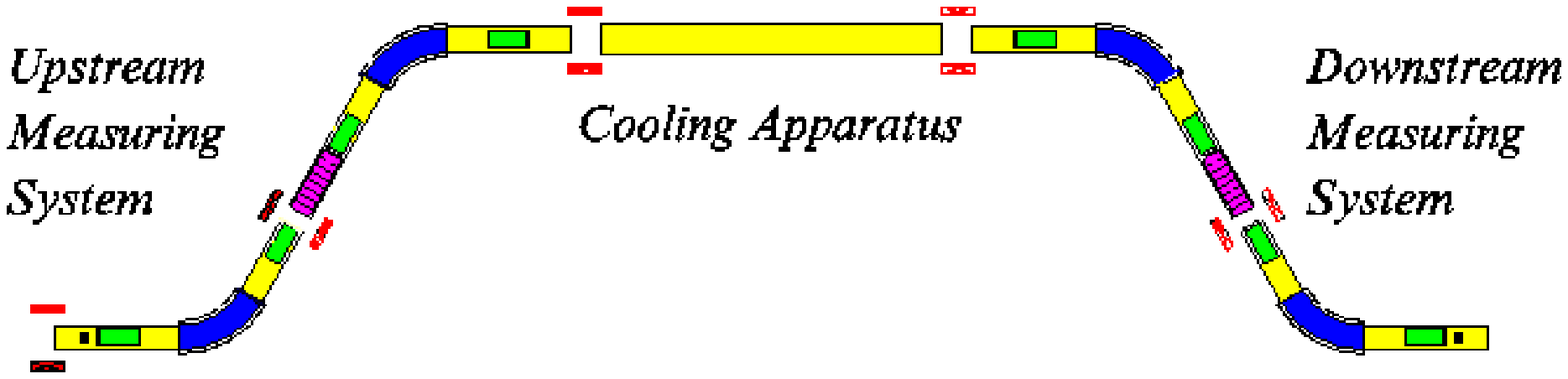}}
\centerline{\epsfxsize=\textwidth\epsffile{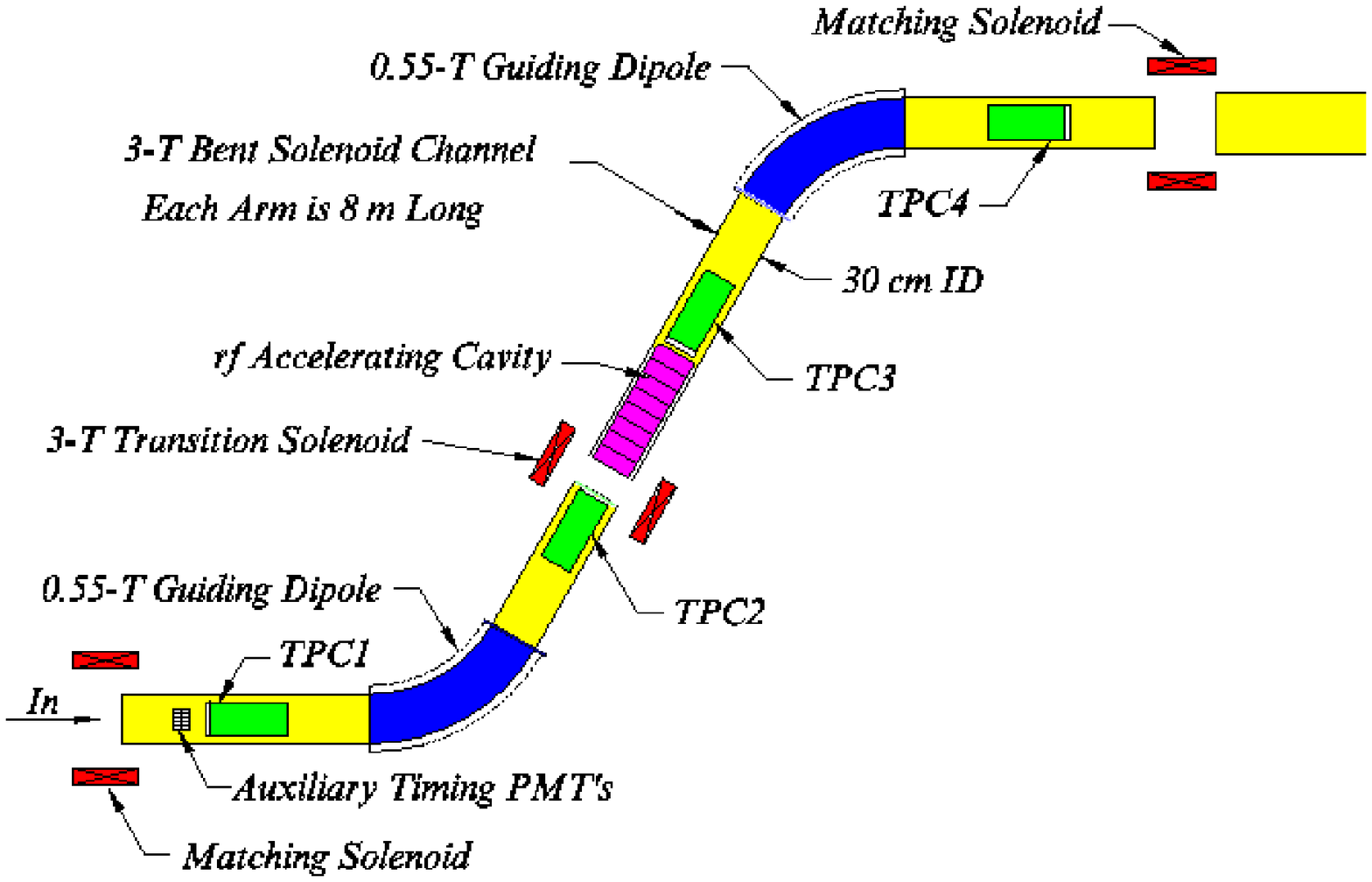}}
\caption{MUCOOL apparatus as proposed. Each muon is measured individually and
the effect of the cooling apparatus reconstructed off-line by combining
individual muons into a ``virtual bunch\@." The muon position and vector momentum
are measured before and after the cooling apparatus using low-pressure TPCs and
bent solenoids. A second momentum measurement on each side allows precision
timing to $\approx$10\,ps by determining the time-dependent momentum kick
imparted to each muon by an RF accelerating module.\label{fig:mucool}}
\end{figure}

While the physics underlying ionization cooling is reasonably well understood,
ionization cooling has yet to be demonstrated in practice. To demonstrate
feasibility and establish performance, a muon cooling experiment
(MUCOOL)~\cite{MUCOOL} has been proposed to Fermilab (Fig.~\ref{fig:mucool}).
With the increasing interest in the possibility of a muon-storage-ring neutrino
factory, over the last year the activities of the NFMCC have undergone a change
of emphasis, and the focus of the proposed experiment is now turning from the
late stages of cooling to the initial stages.  We now envisage a cooling test
facility that will be staged so as to demonstrate initial-stage cooling (such
as will be needed for a neutrino factory) first, with tests of late cooling
stages (needed for a muon collider) coming later. The lower RF frequency used
in early cooling stages relaxes requirements for timing-measurement resolution,
thus simpler measurement approaches than indicated in Fig.~\ref{fig:mucool} are
now envisaged.

\section{Acceleration}

Rapid acceleration to the collider beam energy is needed to avoid excessive
decay losses. This can be accomplished in a series of linacs and ``racetrack"
recirculating linear accelerators (RLAs) such as have been developed at
Jefferson Laboratory. Several scenarios have been considered. As an
example~\cite{PJK}, linacs at 175 and 350\,MHz can be employed to raise the
energy of the cooled muon beam to 2\,GeV, after which it is accelerated in a
first 4-turn RLA to 8\,GeV and, in a second, to 30\,GeV (a possible
storage-ring energy for a neutrino factory). Other scenarios 
include RLAs with larger numbers of turns, ``dogbone-geometry"
RLAs~\cite{CERN-study}, and
fixed-field alternating-gradient~\cite{FFAG} (FFAG) accelerators. At sufficiently high
energy (above a few hundred GeV), the muon lifetime becomes long enough that
ramped ``rapid-accelerating" synchrotrons may be used~\cite{detail}.

\section{Collider Scenarios}

\begin{figure}
\centerline{\epsfxsize=\textwidth\epsffile{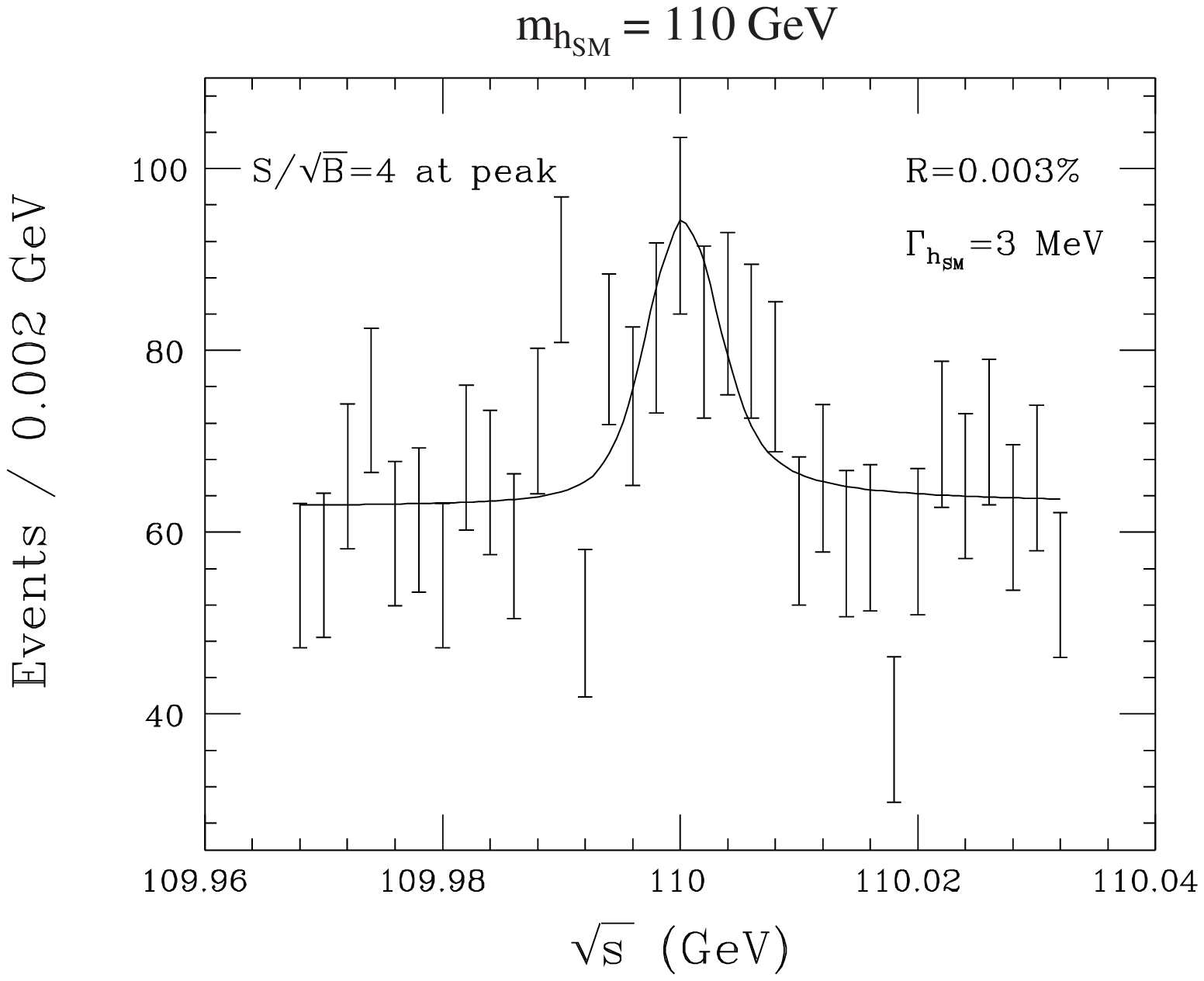}}
\caption{Simulated rate of 
$b{\bar b}$ events for a beam-energy scan in the vicinity of a 
Standard Model
Higgs of 110\,GeV mass, representing about 1 year of running at 
${L}=10^{31}$\,cm$^{-2}$s$^{-1}$ with $\Delta p/p=0.003$\%, and determining the
Higgs mass to $\sim1\,$MeV. 
\label{fig:Higgs}}
\end{figure}

Collider scenarios have been considered at three energies,
$\sqrt{s}=0.1$, 0.4, and 3\,TeV (see Table~\ref{tab:params}). Three variants
of the 0.1\,TeV  (``Higgs Factory") machine have been worked out, covering a
range of momentum spread. While reducing the momentum spread also reduces the
luminosity, given the narrow width expected for the Higgs, the event rate and
Higgs precision are optimized at the narrowest momentum spread
(Fig.~\ref{fig:Higgs})~\cite{Bargeretal}. The 0.4 and 3\,TeV scenarios are
aimed respectively at precision top studies and at searches in a mass regime
beyond the reach of the LHC~\cite{detail}.

\begin{table}[thb!]
\centering  
\caption[Baseline parameters for high- and low-energy muon colliders. ]
{Baseline parameters for high- and low-energy muon colliders.
Higgs/year assumes a cross section $\sigma=5\times 10^4$\,fb and a Higgs width
$\Gamma=2.7$\,MeV (1~year = $10^7$\,s).}
\label{tab:params}
\begin{tabular}{lccccc}
\hline
C.M. energy (TeV)  & 3 & 0.4 &
\multicolumn{3}{c}{0.1 }  \\
$p$ energy (GeV)       &  16  & 16 & \multicolumn{3}{c}{16}\\
$p$/bunch        &  $2.5\times 10^{13}$  & $2.5\times 10^{13}$  &
\multicolumn{3}{c}{$5\times 10^{13}$  }  \\  
Bunches/fill            & 4 & 4 & \multicolumn{3}{c}{2 }  \\
Rep.~rate (Hz)     &  15 & 15 & \multicolumn{3}{c}{15 }  \\
$p$ power (MW)         &  4   & 4 & \multicolumn{3}{c}{4}  \\
$\mu$/bunch   & $2\times 10^{12}$ & $2\times 10^{12}$ &
\multicolumn{3}{c}{$4\times 10^{12}$ }  \\
$\mu$ power (MW)     &  28 & 4 & \multicolumn{3}{c}{ 1 }  \\
Wall power (MW)    &    204 & 120  & \multicolumn{3}{c}{
81 }  \\
Collider circum. (m)          &  6000 & 1000 & \multicolumn{3}{c}{350 }  \\
Avg.\ bending field (T)       & 5.2 & 4.7 &\multicolumn{3}{c}{3 }  \\
r.m.s.\ ${\Delta p/p}$ (\%)          & 0.16 & 0.14 & 0.12 & 0.01& 0.003 \\
6D $\epsilon_{6,N}$ ($(\pi \textrm{m})^3$) &$1.7\times 10^{-10}$&$1.7\times
10^{-10}$&$1.7\times 10^{-10}$&$1.7\times 10^{-10}$&$1.7\times 10^{-10}$\\
r.m.s.\ $\epsilon_n$ ($\pi$ mm-mr)     &  50 & 50 & 85 & 195 & 290\\
$\beta^*$ (cm)          & 0.3 & 2.6 & 4.1 &  9.4 & 14.1\\
$\sigma_z$ (cm)          & 0.3 & 2.6 & 4.1 &  9.4 & 14.1 \\
$\sigma_r$ spot ($\mu$m)     & 3.2 & 26 & 86 & 196 & 294\\
$\sigma_{\theta}$ IP (mr)    & 1.1 & 1.0 & 2.1 & 2.1 & 2.1\\
Tune shift             &0.044 &0.044 & 0.051 &0.022 & 0.015\\
$n_{\rm turns}$ (effective)     &  785 & 700 & 450 & 450 & 450 \\
 Luminosity (cm$^{-2}$s$^{-1}$) & $7\times 10^{34}$ & $10^{33}$ 
&
$1.2\times 10^{32}$ & $2.2\times 10^{31}$& $10^{31}$ \\
Higgs/year     &  & & $1.9\times 10^3$ & $4\times 10^3$ & $3.9\times 10^3$ \\
\hline
\end{tabular}
\end{table}

\section{Collider Detector}

\begin{figure} 
\centerline{\scalebox{0.35}{\includegraphics{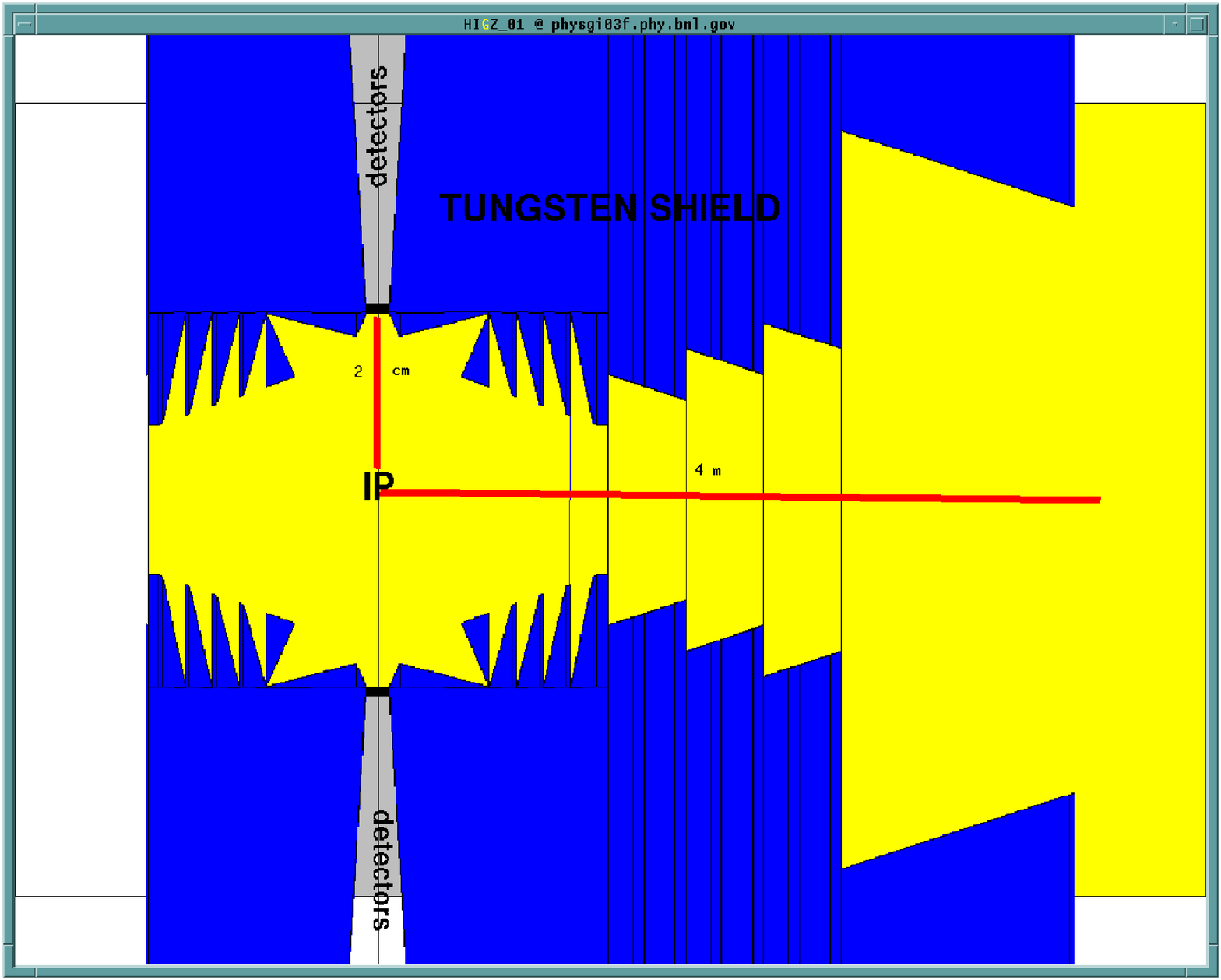}}}
\caption{Possible arrangement of tungsten shielding near the interaction point
(IP), designed so that the detector does not ``see" any surface
hit by decay electrons. 
Scale is indicated by bars extending outwards from the IP: vertical bar
is 2\,cm in length, horizontal bar 4\,m. \label{fig:tungsten}} 
\end{figure}

A ``strawman" detector has been simulated in Geant~\cite{Sitges}. To cope with
high background rates from muon decay within the storage ring, pixel detectors
are employed near the beamline, and an extensive series of tungsten shields are
deployed around the interaction point (Fig.~\ref{fig:tungsten}). At $r=5$\,cm,
pixels of dimensions $60\times150\,\mu$m$^2$ have estimated occupancies below
1\%. Pattern recognition still needs study, but these occupancies are
encouraging.

\section{Neutrino Factory} \label{sec:nufac}

Recently the possibility of a neutrino factory based on a muon storage ring
has received much attention~\cite{Geer,CERN-study,NF,DeRujula,NF2}. Such a facility
could help unravel the mystery of neutrino mixing by providing neutrino beams
of unprecedented intensity, brilliance, and purity.  It could also serve as
a stepping-stone to a muon collider. A neutrino factory should be easier and
cheaper to construct and operate than a muon collider since it requires less
cooling and less intense muon bunches. This follows straightforwardly from
the fact that collider luminosity, being proportional to the square of the
bunch intensity and inversely proportional to the transverse bunch size,
requires as small and intense a bunch as possible, while the sensitivity of
neutrino experiments is determined simply by the time-integrated flux. In
practice this eases the six-dimensional cooling required by a factor
$\sim10^4$, possibly obviating the need for longitudinal-transverse emittance
exchange. 

A muon collider contains various sources of intense neutrino beams, for
example muon decays within the straight sections of the muon accelerators and
collider storage ring. As mentioned above, for sufficiently high muon
energies these ``parasitic" neutrino beams may limit feasibility due to
radiation-safety concerns. Such beams could be used to advance both
``conventional" neutrino physics (structure functions, $\sin^2{\theta_W}$,
etc.) and neutrino-oscillation studies. At high muon energies the flux of
neutrinos is sufficient for useful detection rates on the other side of the
earth. Options that have been discussed include beams aimed from Brookhaven,
CERN, Fermilab,  or KEK to Soudan, SLAC, and Gran Sasso, giving baselines
ranging from 700 to 7000\,km for neutrino-oscillation searches.
Other options are also conceivable, for example use of a new detector at some
suitable location.

Unlike a muon collider, in dedicated-neutrino-factory scenarios the muon
storage ring would be designed to maximize the fraction of muons decaying in
straight sections, leading to an oblong ``racetrack" geometry. Other
geometries have also been considered, e.g.\ a triangular or ``bowtie-shaped"
ring that could aim two neutrino beams simultaneously at two different remote
detectors. The bowtie geometry has the virtue of preserving polarization,
since it bends equally in both directions.

Table~\ref{tab:NF} (from Ref.~\cite{NF2}) exemplifies the physics reach that
may be achievable assuming $2\times10^{20}$ muon decays in a storage-ring
straight section ($\approx$1 year of neutrino-factory running) pointed at a
10\,kiloton detector.  Unlike conventional meson-decay neutrino beams, which
are dominantly $\nu_\mu$ but with some $\nu_e$ contamination, neutrino beams
from a stored $\mu^-$ beam are 50\% $\nu_\mu$ and 50\% $\bar\nu_e$, while
those from $\mu^+$ are 50\% $\bar\nu_\mu$ and 50\% $\nu_e$.  The availability
of intense high-energy electron-neutrino beams makes possible tau- and
muon-neutrino appearance experiments. The use of both muon polarities allows
investigation of matter effects in neutrino oscillation. The phenomenology of
three-flavor neutrino oscillation (as required if at least two of the three
observed effects~\cite{ATM,Solar,LSND} are conclusively established) is quite
complex~\cite{DeRujula,NF2}. No other proposed facility has comparable power
to pin down the details of mixing among three neutrino flavors.

\begin{table}
\centering
\caption{Summary of sensitivity versus baseline and
stored muon energy (from Ref.~\protect\cite{NF2}).
(``Appearance" refers to appearance of tau- or muon-neutrinos, signaled by
detection of ``wrong-sign" muons.)\label{tab:NF}}
\begin{tabular}{cccccc}
\hline
& & \multicolumn{2}{c}{Survival}& \multicolumn{2}{c}{Appearance}\\
\cline{3-4} \cline{5-6}\\[-2ex]   
& &sin$^22\theta_{23}$&$\delta
m^2_{32}$&sin$^22\theta_{13}$&sin$^22\theta_{13}$ \\
& &statistical&statistical& 10~Event&3$\sigma$ sign \\
$L$ (km)& $E_\mu$ (GeV)&precision&precision&sensitivity&$\delta m^2$ \\
\hline
732&  10&7.6\% &6.7\%&0.002 &$>$0.1 \\
732&  30&14\%  &8.9\%&0.0005&0.1   \\
732&  50&17\%  &12\%&0.0003 &$>0.1$ \\
2800& 10&1.1\% &2.4\%&0.008 &0.1   \\
2800& 30&2.0\% &3.2\%&0.0007&0.005 \\
2800& 50&1.8\% &4.9\%&0.0004&0.003 \\
7332& 10&13\%  &6.3\%&0.02  &$>$0.1 \\
7332& 30&0.57\%&1.2\%&0.001 &0.04  \\
7332& 50&0.64\%&1.4\%&0.002 &0.02  \\
\hline
\end{tabular}
\end{table}

\section{Conclusions}

The prospect of a high-luminosity $\mu^+\mu^-$ collider, once entirely 
speculative, has by dint of much work and study now been brought into the
realm of possibility. While a collider is still a question for the long term
(post-2010), ideas (most notably a neutrino factory) spun off from this
effort may have a substantial impact on high-energy physics in the coming
decade.


\begin{thebibliography}{99}

\bibitem{MuColl} 
For a list of Neutrino Factory and Muon Collider Collaboration members, see 
http://pubweb.bnl.gov/people/gallardo/collaboration-list.ps.

\bibitem{Bargeretal}
V. Barger, M. Berger, J. Gunion, and T. Han, Phys.\ Rep.\ {\bf 286}, 1 (1997).

\bibitem{Eichten}
See for example 
R. Casalbuoni {\it et al.}, Phys.\ Lett.\ B {\bf 285}, 103 (1992);
C. T. Hill, Phys.\ Lett.\ B {\bf 345}, 483 (1995);
K. Lane and E. Eichten, Phys.\ Lett.\ B {\bf 352}, 382 (1995);
K. Lane, Phys.\ Rev.\ D {\bf 54}, 2204 (1996);
E. Eichten {\it et al.}, Phys.\ Rev.\ Lett.\ {\bf 80}, 5489 (1998).

\bibitem{Palmer-Gallardo}
R. B. Palmer and J. C. Gallardo, in
{\sl Proc.\ XXVIII Int.\ Conf.\ on 
High Energy Physics}, Z. Ajduk and A. K. Wroblewski, eds., 
World Scientific, Singapore (1997), p.~435.

\bibitem{Front-End}
See {\bf Workshop on
Physics at the First Muon Collider and Front-End of a Muon Collider},
S. Geer and R. Raja, eds., AIP Conf.\
Proc.\ {\bf 435} (American Institute of Physics, New York, 1998).

\bibitem{Status-Report}
C. Ankenbrandt {\it et al.}, Phys.\ Rev.\ ST Accel.\ Beams {\bf 2}, 
081001, 1--73 (1999); 
for a more concise summary see~\protect{\cite{Sitges}}.

\bibitem{stored-muon}
See for example
G. Charpak, L. M. Lederman, J. C. Sens, A. Zichichi, Nuovo Cim.\
{\bf 17}, 288 (1960); 
A. C. Melissinos (unpublished, 1960; scanned copy at
http://www.hep.princeton.edu/mumu/physics/index.html); 
J. Tinlot and D. Green, UR-875-76 (1965).

\bibitem{Tikhonin}
F. F. Tikhonin, JINR Report P2-4120 
(Dubna, 1968).

\bibitem{ONeill}
G. K. O'Neill, Phys.\ Rev.\ {\bf 102}, 1418 (1956); 
A. A.
Kolomensky, Sov.\ Atomic Energy {\bf 19}, 1511 (1965); 
G. I. Budker and A. N. Skrinsky,
Sov.\ Phys.\ Usp.\ {\bf 21}, 277 (1978); 
D. Neuffer, FNAL Report FN-319 (1979); 
A. N. Skrinsky
and V. V. Parkhomchuk, Sov.\ J. Nucl.\ Phys.\ {\bf 12}, 223 (1981); 
D. Neuffer,
Part.\ Acc.\ {\bf 14}, 75 (1983); 
E. A. Perevedentsev and A. N. Skrinsky, in 
{\sl Proc.\ 12th Int.\ Conf.\ on High 
Energy 
Accelerators}, F.~T.~Cole and R. Donaldson, eds.\ (1983), p.~485.

\bibitem{Lichtenberg}
D. B.
Lichtenberg, P. Stehle, and K. R. Symon, MURA-126
(unpublished, 1956; scanned copy at
http://www.hep.princeton.edu/mumu/physics/index.html).

\bibitem{Neuffer2}
D. Neuffer, in {\bf Advanced Accelerator Concepts}, F. E. Mills, ed., AIP 
Conf.\ Proc.\ {\bf 156} (American Institute of Physics, New York, 1987), p.~201.

\bibitem{targetry}
J. Alessi {\it et al.}, Brookhaven AGS Proposal 951, Sept.\ 28, 1998,
http://www.hep.princeton.edu/mumu/target/targetprop.ps.

\bibitem{SNS}
See http://www.ornl.gov/sns/sect6.htm.

\bibitem{BNL-source}
T. Roser, in {\bf Workshop on Space Charge Physics in High Intensity 
Hadron Rings}, A. U. Luccio and W. T. Weng, eds., 
AIP Conf.\ Proc.\ {\bf 448} (American Institute of Physics, 
New York, 1998), p.\ 135.

\bibitem{FNAL-source}
S. D. Holmes, Fermilab Report Fermilab-TM-2021 (1997); 
W. Chou, {\sl Proc.\ 1999 Particle Accelerator Conference}, A. Luccio and W.
MacKay, eds.\ (IEEE, New York, 1999), p.~3285.

\bibitem{BNL-SNS}
Brookhaven National Laboratory Report BNL-60678 (1994).

\bibitem{Blondel} 
A. Blondel, 
in  ``Prospective Study of Muon Storage Rings at
CERN," B. Autin, A. Blondel, and J. Ellis, eds., CERN 99-02, ECFA 99-197
(1999), p.\ 51.

\bibitem{Geer}
S. Geer, Phys.\ Rev.\ D {\bf 57}, 6989 (1998).

\bibitem{Raja}
R. Raja and A. Tollestrup, Phys.\ Rev.\ D {\bf 58}, 13005 (1998).

\bibitem{Fernow}
R. C. Fernow and J. C. Gallardo, Phys.\ Rev.\ E {\bf 52}, 1039 (1995).

\bibitem{PDG-accel}
K. Desler and D. A. Edwards, ``Accelerator Physics of Colliders,"
in C. Caso {\it et al.}, Eur.\ Phys.\ J. {\bf C3} (1998) 1.

\bibitem{PDG}
C. Caso {\it et al.}, {\it ibid.}

\bibitem{2mSoSo}
J. C. Gallardo {\it et al.}, 
{\sl Proc.\ 1999 Particle Accelerator Conference},
{\it op.\ cit.}, p.~3032.

\bibitem{Fernowetal}
R. C. Fernow, J. C. Gallardo, H. G. Kirk, and R. B. Palmer
http://pubweb.bnl.gov/people/fernow/reports/asol.ps (to be published).

\bibitem{Mark-Beise}
J. W. Mark, SLAC-PUB-3169 (1984) and references therein; 
E. J. Beise {\it et al.}, Nucl.\ Instrum.\ Meth.\ {\bf A378}, 383 
(1996).

\bibitem{Margaziotis-E158}
D. J. Margaziotis, in {\sl Proc.\ CEBAF Summer 1992 Workshop}, F. Gross and 
R. Holt, eds., AIP Conf.\ Proc.\ {\bf
269} (American Institute of Physics, New York, 1993), p.~531; 
R. Carr {\it et al.}, SLAC-Proposal-E-158, July 1997.

\bibitem{Scanlan:1996qp}
R. M.~Scanlan {\it et al.},
Nucl.\ Instrum.\ Meth.\ {\bf A380}, 544 (1996);
M. Green, 
private
communication; 
J. Miller, 
private communication.

\bibitem{Snowmass}
R. B. Palmer, A. Sessler, and A. Tollestrup, in {\sl Proc.\ 1996 DPF/DPB Summer 
Study on High-Energy Physics}, D. G. Cassel, L. T. Gennari, and R. H. Siemann, 
eds.\ (Stanford Linear Accelerator Center, Menlo Park, CA, 1997), 
p.~203; 
The $\mu^+\mu^-$ Collider Collaboration, Report No.\ BNL-52503, 
Fermilab-Conf-96-092, LBNL-38946 (1996).

\bibitem{Corlett}
J. N. Corlett {\it et al.}, 
{\sl Proc.\ 1999 Particle Accelerator Conference}, {\it op.\ cit.}, p.~3149.

\bibitem{MUCOOL}
C. N. Ankenbrandt {\it et al.}, 
Fermilab Proposal 904 (April 15, 1998).

\bibitem{PJK}
R. B. Palmer, C. Johnson, and E. Keil, 
 BNL-6971, CERN SL/99-070 AP, to appear in {\sl Proc.\
Lyon Neutrino Factory Workshop} (1999).

\bibitem{CERN-study}
``Prospective Study of Muon Storage Rings
  at CERN," B. Autin, A. Blondel, and J. Ellis, eds., CERN 99-02, ECFA 99-197 (1999); 
Nu-Fact'99, Lyon, 5--9 July 1999.

\bibitem{FFAG}
K. R. Symon {\it et al.}, Phys.\ Rev.\ {\bf 103}, 6 (1956).

\bibitem{detail}
Much more detail on acceleration and storage-ring designs is given in 
Ref.~\protect\cite{Status-Report}.

\bibitem{Sitges}
R. Raja, Fermilab-Conf-99/329, to appear in 
{\it Proc.\ Worldwide Study on Physics and Experiments with Future Linear 
Colliders}, Sitges (Barcelona), Spain, April 28--May 5, 1999.

\bibitem{NF} 
S. Geer, in {\bf Workshop on
Physics at the First Muon Collider and Front-End of a Muon Collider}, 
{\it op.\ cit.}, p.~384; 
V. Barger, S. Geer, and K. Whisnant, Fermilab-Pub-99-187-T (1999); 
{\sl Workshop on the Potential for Neutrino Physics at Future Muon Colliders},
Brookhaven National Laboratory,  Aug.\ 13--14, 1998, 
http://pubweb.bnl.gov/people/bking/nushop/workshop.html.

\bibitem{DeRujula}
A. De Rujula, M. B. Gavela, and P. Hernandez, Nucl.\ Phys.\ {\bf B547}, 21
(1999).

\bibitem{NF2}
V. Barger, S. Geer, R. Raja, and K. Whisnant, Fermilab-Pub-99-341-T (1999), and
references therein.

\bibitem{ATM}
Super-Kamiokande collab.,
Y. Fukuda et al., Phys.\ Lett.\ B {\bf 433}, 9 (1998);
Phys.\ Lett.\ B {\bf 436}, 33 (1998);
Phys.\ Rev.\ Lett.\ {\bf 81}, 1562 (1998);
Phys.\ Rev.\ Lett.\ {\bf 82}, 2644 (1999);
Kamiokande collab., K. S. Hirata {\it et al.}, Phys.\ Lett.\ B
{\bf 280}, 146 (1992); Y. Fukuda {\it et al.}, Phys.\ Lett.\ B
{\bf 335}, 237 (1994);
IMB collab., R. Becker-Szendy {\it et al.}, Nucl.\ Phys.\ Proc.\ 
Suppl.\ {\bf 38B}, 331 (1995);
Soudan-2 collab., W. W. M. Allison {\it et al.}, Phys.\ Lett.\ B {\bf
391}, 491 (1997);
MACRO collab., M. Ambrosio {\it et al.},
Phys.\ Lett.\ B {\bf 434}, 451 (1998).

\bibitem{Solar}
B. T. Cleveland {\it et al.}, Nucl.\ Phys.\ B Proc.\ Suppl.\ {\bf 38}, 47
(1995);
GALLEX collab., W. Hampel {\it et al.}, Phys.\ Lett.\ B {\bf 388},
384 (1996);
SAGE collab., J. N. Abdurashitov {\it et al.}, Phys.\ Rev.\ Lett.\ 
{\bf 77}, 4708 (1996);
Kamiokande collab., Y. Fukuda {\it et al.}, Phys.\ Rev.\ Lett.\ 
{\bf 77}, 1683 (1996);
Super-Kamiokande collab., Y. Fukuda {\it et al.},
Phys.\ Rev.\ Lett.\ {\bf 82}, 2430 (1999);
Phys.\ Rev.\ Lett.\ {\bf 82}, 1810 (1999);
J. N. Bahcall, S. Basu, and M. H. Pinsonneault, Phys.\ Lett.\ B
{\bf 433}, 1 (1998), and references therein.

\bibitem{LSND}
C. Athanassopoulos {\it et al.} (LSND collab.),
Phys.\ Rev.\ Lett.\ {\bf 77}, 3082 (1996);
Phys.\ Rev.\ Lett.\ {\bf 81}, 1774 (1998).

\end{thebibliography}
\end{document}